\documentclass[12pt,a4paper]{article}

\usepackage{amsthm,amsmath,natbib,amssymb,amsfonts,bm, mathtools}
\usepackage{graphicx}
\usepackage{placeins}
\usepackage{epstopdf}
\usepackage{secdot}
\usepackage{longtable}
\usepackage{xcolor}
\usepackage{listings}
\usepackage{subfig}
\usepackage{arydshln}
\usepackage{bm}
\usepackage{upgreek}
\usepackage{url}
\usepackage{enumitem}

\newtheorem*{theorem*}{Theorem}

\theoremstyle{definition}

\theoremstyle{remark}

\DeclareMathAlphabet{\mathpzc}{OT1}{pzc}{m}{it}

\newcommand{\E}{\text{E}}

\makeatletter
\newcommand{\vast}{\bBigg@{4}}
\newcommand{\Vast}{\bBigg@{5}}
\makeatother

\begin{document}
	
\baselineskip = 20pt	
	
\begin{center}	
\textbf{\Large Custom-made Gauss quadrature for statisticians}
\end{center}	

\medskip

\begin{center}	
\textbf{\large Paul Kabaila$^*$}
\end{center}	


\begin{center}	
	\textsl{Department of Mathematical and Physical Sciences, 
	La Trobe University, Australia}
\end{center}	
	
		\vspace{1cm}
		
		\begin{center}
		    \textbf{Abstract}
		\end{center}
	
\noindent 		The theory and computational methods for custom-made Gauss quadrature have been  
described in Gautschi's 2004 monograph. Gautschi has also provided \texttt{Fortran} and \texttt{MATLAB} code for the implementation and illustration of these methods. We have written an \texttt{R} package, implemented in the high-precision arithmetic provided by the \texttt{R} package \texttt{Rmpfr}, that uses a moment-based method via moment determinants to compute a Gauss quadrature rule, with up to 33 nodes, provided that the moments can be computed to arbitrary precision using the standard mathematical functions provided by the \texttt{Rmpfr} package.  
  Our hope is that the provision of our free \texttt{R} package and the numerical results that we present will encourage other statisticians to also consider the custom-made construction of Gauss quadrature rules.

\vspace{8cm}

\noindent * Postal address: Department of Mathematical and Physical Sciences, 
La Trobe University, Bundoora Victoria 3086, Australia.
Email: P.Kabaila@latrobe.edu.au	 

	\newpage

\section{Introduction}

In statistical applications, we commonly wish to evaluate 
\begin{equation}
\label{eqn_integral_to_be_evaluated}	
\int_{-\infty}^{\infty}	g(x) \, f(x) \, dx,
\end{equation}
where $f$ is a specified nonnegative integrable weight function on $\mathbb{R}$. In the particular case that $f$ is the probability density function (pdf) of a random variable $X$, \eqref{eqn_integral_to_be_evaluated} is equal to
$\E\big(g(X)\big)$. The Gauss quadrature approximation to this integral has the form
\begin{equation*}
\sum_{i = 1}^n \lambda_{i} \, g(\tau_{i}),
\end{equation*}
where $\tau_1, \dots, \tau_n$ are called the nodes and 
$\lambda_1, \dots, \lambda_n$ are called the corresponding weights.
The dependence of these nodes and weights on $n$ is implicit. This approximation is exact whenever $g$ is a polynomial of degree less than or equal to $2n - 1$.

If $f$ takes a particular form that leads to Gauss quadrature rules with nodes that are the roots of 
classical orthogonal polynomials of a continuous variable (such as Legendre, Hermite and Laguerre polynomials) then these rules are readily accessible to statisticians via \texttt{R} packages such as \texttt{statmod} (\citeauthor{statmod},
\citeyear{statmod}). If, however, $f$ does not take one of these particular forms then the Gauss quadrature rule needs to be custom-made. This can be done by computing orthogonal polynomials with respect to the measure specified by the weight function $f$, right up to 
$\pi_n(\, \cdot \, )$ the monic orthogonal polynomial of degree $n$ with respect to this measure. The nodes 
$\tau_1, \dots, \tau_n$ are the roots of $\pi_n(\, \cdot \, )$.

An important problem in statistics (see \citeauthor{KabailaRanathunga2021},  \citeyear{KabailaRanathunga2021}, for references) is to rapidly compute an accurate approximation to the integral of the form 
\eqref{eqn_integral_to_be_evaluated} when the weight function $f$ is as specified in the following example. For this example, the Gauss quadrature rule needs to be custom-made. 

\smallskip

\noindent \textbf{Example (scaled chi pdf)}:
Suppose that $f$ is the probability density function (pdf) of a random variable $X$ with the same probability distribution as $R / m^{1/2}$, where $R$ has a $\chi_m$ distribution (i.e. $R^2$ has a $\chi_m^2$ distribution). The name given to $m$ is the degrees of freedom for the $\chi_m$ distribution. This pdf is given by 
\begin{equation}
	\label{eqn_pdf_of_Example}
	\begin{split}
	f(x) =
	\begin{cases}
		\dfrac{m^{m/2}}{\Gamma(m/2) \, 2^{(m/2) - 1}} \; x^{m-1} \, \exp\big(- m \, x^2 /2\big) &\text{for } x > 0
		\\
		0 &\text{otherwise}.
	\end{cases}	
\end{split}
\end{equation}

\smallskip

The theory and computational methods for custom-made Gauss quadrature rules was
described by \cite{Gautschi1994}. He provided a suite of \texttt{Fortran} code for the implementation and illustration of these methods. 
Physicists have used this suite of \texttt{Fortran} code to provide custom-made Gauss quadrature rules, see e.g. \cite{GanderKarp2001}.
An exhaustive description of this theory and these computational methods was provided by \cite{Gautschi2004}. He provided a suite of \texttt{MATLAB} code for the implementation and illustration of these methods.
One of these methods (with some small modification) has also been implemented in \texttt{C++} code by computational biologists, see \cite{FernandesAtchley2006}. \cite{FukudaEtAl2005} used both the symbolic algebra and high-precision arithmetic capabilities of \texttt{Mathematica} to compute custom-made Gauss quadrature rules.

We have implemented the moment-based method via moment determinants described in subsection 2.1.1 of \cite{Gautschi2004}
for computing custom-made Gauss quadrature rules in the \texttt{R} 
package \texttt{custom.gauss.quad}.
This method is severely ill-conditioned
(Gautschi, \citeyear{Gautschi1968}, \citeyear{Gautschi1983}, 
\citeyear{Gautschi1994}, \citeyear{Gautschi2004}). However, for Gauss quadrature with up to 33 nodes, its severe limitations can be overcome by the use of high-precision arithmetic. 
We use the high-precision arithmetic provided via the \texttt{R} package \texttt{Rmpfr} (\citeauthor{Rmpfr}, \citeyear{Rmpfr}) which ultimately provides an interface to the GNU Multiple Precision arithmetic library (GMP). An advantage of our \texttt{R} package is that \texttt{R} is an open-source free software environment. Furthermore, statisticians are generally familiar with \texttt{R} packages.

We have sought to minimize our code-writing effort by using 
the off-the-shelf functions provided by the 
 \texttt{R} 
package \texttt{Rmpfr} as much as possible. 
Our \texttt{R} package
can be used for the computation of custom-made Gauss quadrature rules, with up to 33 nodes, for nonnegative weight functions $f$, 
provided that there is a
formula for the $r$'th moment
\begin{equation}
\label{eq_moments_as_integrals}	
\mu_r
= \int_{-\infty}^{\infty}	x^r \, f(x) \, dx
\end{equation}
for all nonnegative integers $r$,
in terms of the mathematical functions that can be computed to arbitrary precision using the \texttt{R} package \texttt{Rmpfr}.
In the particular case that $f$ is the pdf of a random variable $X$, \eqref{eq_moments_as_integrals} is equal to
$\E\big(X^r\big)$.
Our hope is that the provision of our \texttt{R} package and the numerical results that we present will encourage other statisticians to also consider the custom-made construction of Gauss quadrature rules, 
using not only our \texttt{R} package but also one of the other methods described by \cite{Gautschi2004} that are applicable for  
a larger number of nodes,
for the numerical integration problems that they encounter.

\section{Moment-based method via moment \newline determinants}
\label{sect_moment_based_method}

Our \texttt{R} package \texttt{custom.gauss.quad}  computes the custom-made Gauss quadrature rule, 
where the first step is a moment-based method using moment determinants. Our 
code uses the high-precision arithmetic provided by the \texttt{R} package \texttt{Rmpfr} throughout. 
This code is applicable whenever there is a
formula for the moments
in terms of the mathematical functions that can be computed to arbitrary precision using this package. 
In this section, we suppose that the number of nodes $n$ is specified. 
 For given number \texttt{b} of  bits of precision, the steps carried out by our code are described in Appendix A.  Next, we turn our attention to the problem of seeking some assurance that the chosen number \texttt{b} of  bits of precision is sufficiently large for our purposes.

Real numbers in \texttt{R} are stored in double precision, with a precision of 53 bits, which is approximately 16 decimal digits. 
Ultimately, the nodes and weights that are computed by our \texttt{R} package 
\texttt{custom.gauss.quad} are converted to 
this double precision form using the \texttt{asNumeric} function from the \texttt{R} package \texttt{Rmpfr}. These nodes and weights are then available for further extensive double precision computations in 
\texttt{R}.

Let $\tau_1(\texttt{b}), \dots, \tau_n(\texttt{b})$
and $\lambda_1(\texttt{b}), \dots, \lambda_n(\texttt{b})$
denote the high-precision nodes and weights, respectively, computed  using our \texttt{R} code that implements the steps described in Appendix A, when 
$\texttt{b}$ bits of precision are used.
We compute these nodes and weights for an increasing sequence $\texttt{b}_1, \dots, \texttt{b}_M$ of values of $\texttt{b}$, where
$\texttt{b}_j = \texttt{b}_1 + 34 \, (j-1)$ for $j = 2, \dots, M$.
The means that the increase in precision as we go from 
$\texttt{b}_j$ to $\texttt{b}_{j+1}$ bits is roughly 10 decimal digits.
In this paper, we have chosen $M = 5$ and 
$\texttt{b}_1$ to be the smallest integer not less than 
$60 + 6.5 \, n$. 

We obtain some degree of assurance that the number of bits 
$\texttt{b}_M$ is sufficiently large for our purposes 
as follows. If, after 
conversion to double precision in \texttt{R} using the \texttt{asNumeric} function from the package \texttt{Rmpfr}, 
 the value of $\big(\tau_1(\texttt{b}_j), \dots, \tau_n(\texttt{b}_j)\big)$
is the same for $j = M-1$ and $j = M$ then we define
$L_{\text{nodes}}$ to be the smallest positive integer $k$ such that, after conversion to double precision using the \texttt{asNumeric} function, 
the value of $\big(\tau_1(\texttt{b}_j), \dots, \tau_n(\texttt{b}_j)\big)$
is the same for $j = k, \dots, M$. 
If, after 
conversion to double precision in \texttt{R} using the \texttt{asNumeric} function from the package \texttt{Rmpfr}, 
the value of 
$\big(\lambda_1(\texttt{b}_j), \dots, \lambda_n(\texttt{b}_j)\big)$
is the same for $j = M-1$ and $j = M$ then we define
$L_{\text{weights}}$ to be the smallest positive integer $k$ such that, after conversion to double precision using the \texttt{asNumeric} function, 
the value of $\big(\lambda_1(\texttt{b}_j), \dots, \lambda_n(\texttt{b}_j)\big)$
is the same for $j = k, \dots, M$. 
Also let
\begin{align*}
	d_{\tau}(j) 
	&=	\max \Big( 
	\big| \tau_1(\texttt{b}_{j-1}) - \tau_1(\texttt{b}_{j}) \big|, \dots,
	\big|\tau_n(\texttt{b}_{j-1}) - \tau_n(\texttt{b}_{j}) \big|	\Big)
	\\
	d_{\lambda}(j) 
	&= \sum_{i = 1}^n 
	\big| \lambda_i(\texttt{b}_{j-1}) - \lambda_i(\texttt{b}_{j}) \big|,
\end{align*}
for $j = 2, \dots, M$. The smaller are $L_{\text{nodes}}$ and
$L_{\text{weights}}$ and the more rapidly 
and systematically 
$d_{\tau}(j)$ and 
$d_{\lambda}(j)$ both decrease towards zero as $j$ increases 
($j = 2, \dots, M$), the greater the degree of assurance that 
number of bits 
$\texttt{b}_M$ is sufficiently large for our purposes.

\section{Overview of the \texttt{R} package \texttt{custom.gauss.quad}}
\label{sect_overview_package}

The \texttt{R} package \texttt{custom.gauss.quad} provides the two \texttt{R} functions
\texttt{moments} and \texttt{custom}. The \texttt{R} function \texttt{moments}
computes $\mu_r$, the $r$'th moment, of the specified nonnegative weight function $f$
to an arbitrary number of bits of precision using the  
mathematical functions that can be computed to arbitrary precision via the 
\texttt{R} package \texttt{Rmpfr}. Further details about this function are provided Appendix B. 

The \texttt{R} function \texttt{custom} computes the Gauss quadrature nodes and weights, for number of nodes $n$ and the specified nonnegative weight function $f$, using the moments computed using the \texttt{R} function \texttt{moments}. The inputs and outputs of the \texttt{R} function \texttt{custom} are described in Appendix C.

\section{Numerical results for our Example}

In this section, we illustrate the application of our \texttt{R} package to our Example (scaled chi pdf), for which  
it may be shown that $\mu_r$, the $r$'th moment,  is 
\begin{equation}
\label{eqn_moments_scaled_chi_pdf}
	\left(\frac{2}{m} \right)^{r/2}
	\frac{\Gamma\big((r+m)/2\big)}{\Gamma(m/2)},
\end{equation}
which can be computed to arbitrary precision using the \texttt{R} package \texttt{Rmpfr}. 
We consider both the small value $m = 2$ and the  large value $m = 160$.
 All 
timings are for a computer with Intel i7-3770 CPU $@$ 3.40GHz and 16GB of RAM. In every case, the computed values of 
$\big(d_{\tau}(2), d_{\tau}(3), d_{\tau}(4), d_{\tau}(5)\big)$,
$\big(d_{\lambda}(2),d_{\lambda}(3), d_{\lambda}(4), d_{\lambda}(5)\big)$, $L_{\text{nodes}}$ and
$L_{\text{weights}}$ 
provide us with some reasonable degree of assurance that 
$\texttt{b}_5$ bits of precision is sufficiently large that, after
the nodes and weights are converted to double precision using  
the \texttt{asNumeric} function from the \texttt{R} package \texttt{Rmpfr}, the maximum possible precision (in double precision) is achieved.

\subsection{$\boldsymbol{m=2}$}

Throughout this subsection, we suppose that $m = 2$.

\subsubsection{$\boldsymbol{n = 5}$}

Suppose that $n = 5$ and 
$\big(\texttt{b}_1, \dots, \texttt{b}_5 \big) 
= (93, 127, 161, 195, 229)$. 
For each of these values of \texttt{b}, Steps 1 and 2 (taken together) took roughly 3 seconds, Step 3 took roughly 2 seconds and Step 4 took roughly 27 seconds. These computations resulted in approximate values 
$\big(d_{\tau}(2), d_{\tau}(3), d_{\tau}(4), d_{\tau}(5)\big)
= \big(4.0 \times 10^{-25}, 2.6 \times 10^{-35},
 4.2 \times 10^{-45},  1.7 \times 10^{-55}\big)$ and
$\big(d_{\lambda}(2),d_{\lambda}(3), d_{\lambda}(4), d_{\lambda}(5)\big) = \big(5.7 \times 10^{-25},  3.3 \times 10^{-35}, 5.4 \times 10^{-45},  2.4 \times 10^{-55}\big)$. Also $L_{\text{nodes}} = 1$ and
$L_{\text{weights}} = 1$. 

\subsubsection{$\boldsymbol{n = 17}$}

Suppose that $n = 17$ and 
$\big(\texttt{b}_1, \dots, \texttt{b}_5 \big) 
= (171, 205, 239, 273, 307)$. 
For each of these values of \texttt{b}, Steps 1 and 2 (taken together) took roughly 300 seconds, Step 3 took roughly 10 seconds and Step 4 took roughly 96 seconds. These computations resulted in approximate values 
$\big(d_{\tau}(2), d_{\tau}(3), d_{\tau}(4), d_{\tau}(5)\big)
= \big(7.8 \times 10^{-35}, 7.1 \times 10^{-45},  1.9 \times 10^{-54}, 6.2 \times 10^{-65}\big)$ and
$\big(d_{\lambda}(2),d_{\lambda}(3), d_{\lambda}(4), d_{\lambda}(5)\big) = \big(1.2 \times 10^{-34}, 8.5 \times 10^{-45}, 2.0 \times 10^{-54},  6.4 \times 10^{-65}\big)$. Also $L_{\text{nodes}} = 1$ and
$L_{\text{weights}} = 1$. 

\subsubsection{$\boldsymbol{n = 33}$}

Suppose that $n = 33$ and 
$\big(\texttt{b}_1, \dots, \texttt{b}_5 \big) 
= (275, 309, 343, 377, 411)$. 
For each of these values of \texttt{b}, Steps 1 and 2 (taken together) took roughly 56 minutes, Step 3 took roughly 24 seconds and Step 4 took roughly 188 seconds. These computations resulted in approximate values 
$\big(d_{\tau}(2), d_{\tau}(3), d_{\tau}(4), d_{\tau}(5)\big)
= \big(4.0 \times 10^{-25}, 2.6 \times 10^{-35},  4.2 \times 10^{-45}, 1.7 \times 10^{-55}\big)$ and
$\big(d_{\lambda}(2),d_{\lambda}(3), d_{\lambda}(4), d_{\lambda}(5)\big) = \big(5.7 \times 10^{-25}, 3.3 \times 10^{-35}, 5.4 \times 10^{-45},  2.4 \times 10^{-55}\big)$. Also $L_{\text{nodes}} = 1$ and
$L_{\text{weights}} = 1$.

\subsection{$\boldsymbol{m=160}$}

Throughout this subsection, we suppose that $m = 160$.

\subsubsection{$\boldsymbol{n = 5}$}

Suppose that $n = 5$ and 
$\big(\texttt{b}_1, \dots, \texttt{b}_5 \big) 
= (93, 127, 161, 195, 229)$. 
For each of these values of \texttt{b}, Steps 1 and 2 (taken together) took roughly 3 seconds, Step 3 took roughly 2 seconds and Step 4 took roughly 27 seconds. These computations resulted in approximate values 
$\big(d_{\tau}(2), d_{\tau}(3), d_{\tau}(4), d_{\tau}(5)\big)
= \big(2.8 \times 10^{-16}, 
8.1 \times 10^{-27}, 
4.3 \times 10^{-38}, 
1.5 \times 10^{-46}\big)$ and
$\big(d_{\lambda}(2),d_{\lambda}(3), d_{\lambda}(4), d_{\lambda}(5)\big) = \big(3.3 \times 10^{-15},
9.7 \times 10^{-26},
3.6 \times 10^{-37}, 
1.9 \times 10^{-45}\big)$. Also $L_{\text{nodes}} = 2$ and
$L_{\text{weights}} = 2$. 

\subsubsection{$\boldsymbol{n = 17}$}

Suppose that $n = 17$ and 
$\big(\texttt{b}_1, \dots, \texttt{b}_5 \big) 
= (171, 205, 239, 273, 307)$. 
For each of these values of \texttt{b}, Steps 1 and 2 (taken together) took roughly 260 seconds, Step 3 took roughly 13 seconds and Step 4 took roughly 94 seconds. These computations resulted in approximate values 
$\big(d_{\tau}(2), d_{\tau}(3), d_{\tau}(4), d_{\tau}(5)\big)
= \big(5.1 \times 10^{-14}, 
1.0 \times 10^{-24}, 
7.9 \times 10^{-35}, 
8.5 \times 10^{-46}\big)$ and
$\big(d_{\lambda}(2),d_{\lambda}(3), d_{\lambda}(4), d_{\lambda}(5)\big) 
= \big(7.5 \times 10^{-13},
1.4 \times 10^{-23},
1.2 \times 10^{-33}, 
1.2 \times 10^{-44}\big)$. Also $L_{\text{nodes}} = 2$ and
$L_{\text{weights}} = 2$. 

\subsubsection{$\boldsymbol{n = 33}$}

Suppose that $n = 33$ and 
$\big(\texttt{b}_1, \dots, \texttt{b}_5 \big) 
= (275, 309, 343, 377, 411)$. 
For each of these values of \texttt{b}, Steps 1 and 2 (taken together) took roughly 58 minutes, Step 3 took roughly 37 seconds and Step 4 took roughly 189 seconds. These computations resulted in approximate values 
$\big(d_{\tau}(2), d_{\tau}(3), d_{\tau}(4), d_{\tau}(5)\big)
= \big(2.1 \times 10^{-17}, 
1.8 \times 10^{-27}, 
6.8 \times 10^{-39}, 
1.1 \times 10^{-48}\big)$ and
$\big(d_{\lambda}(2),d_{\lambda}(3), d_{\lambda}(4), d_{\lambda}(5)\big) = \big(2.9 \times 10^{-16},
2.6 \times 10^{-26},
9.1 \times 10^{-38}, 
1.6 \times 10^{-47}\big)$. Also $L_{\text{nodes}} = 2$ and
$L_{\text{weights}} = 2$.


\section{Check of \texttt{R} code by comparing with some known \newline results}

To check our \texttt{R} code, we have compared our computed high-precision results,  with known results for some of the classical recursion coefficients and Gauss quadrature nodes and weights.

\subsection{Computation of classical recursion coefficients $\boldsymbol{\alpha_k}$ and $\boldsymbol{\beta_k}$}

We check our \texttt{R} code for Step 1 of Appendix A, by comparing our computed high-precision results (using $\texttt{b} = 411$  bits of precision) for the recursion coefficients $\alpha_k$ and $\beta_k$ with known results, for $k = 0, 1, 2, \dots , 32$. 

\subsubsection{Hermite}

Consider the Hermite weight function
\begin{equation}
\label{eqn_Hermite_weight_function}
f(x) = \exp(-x^2) \ \text{for all} \  x \in \mathbb{R}.
\end{equation}
According to Example 2.6 of \cite{Gautschi2004},
the $r$'th moment 
\begin{equation}
\label{eqn_moments_Hermite}	
\begin{split}
\mu_r =
\begin{cases}
	0 & \text{for}\ r \ \text{odd}
	\\
	\Gamma \big((r + 1) / 2\big) & \text{for}\ r \ \text{even}.
\end{cases}	
\end{split}	
\end{equation}
According to Table 1.1 of 
\cite{Gautschi2004}, $\alpha_k = 0$ for $k = 0, 1, 2, \dots$,
$\beta_0 = \sqrt{\pi}$ and $\beta_k = k / 2$ for $k = 1, 2, 3, \dots$. The maximum of the absolute values of the differences between the computed recursion coefficients and these known results was approximately $1.9 \times 10^{-110}$. 

\subsubsection{Legendre}

Consider the Legendre weight function 
\begin{equation}
\label{eqn_Legendre_weight_function}
\begin{split}
f(x) =
\begin{cases}
	1 &\text{for} \ \ x \in [-1, 1]
	\\
	0 &\text{otherwise}.
\end{cases}	
\end{split}
\end{equation}
The $r$'th moment 
\begin{equation}
	\label{eqn_moments_Legendre}
\begin{split}
	\mu_r =
	\begin{cases}
		0 & \text{for}\ r \ \text{odd}
		\\
	    2 / (r + 1) & \text{for}\ r \ \text{even}.
	\end{cases}	
\end{split}	
\end{equation}
According to Table 1.1 of 
\cite{Gautschi2004}, $\alpha_k = 0$ for $k = 0, 1, 2, \dots$,
$\beta_0 = 2$ and $\beta_k = 1 / (4 - k^{-2})$ for $k = 1, 2, 3, \dots$.
The maximum of the absolute values of the differences between the computed recursion coefficients and these known results was approximately $1.7 \times 10^{-103}$.  

\subsubsection{Generalized Laguerre}

Consider the Generalized Laguerre weight function 
\begin{equation}
	\label{eqn_Generalized_Laguerre_weight_function}
	f(x) =
	\begin{cases}
		x^{\alpha} \, \exp(-x) &\text{for} \ \ x \in [0, \infty)
		\\
		0 &\text{otherwise},
	\end{cases}	
\end{equation}
where $\alpha > -1$.
The $r$'th moment 
\begin{equation}
	\label{eqn_moments_Generalised_Laguerre}
	\mu_r = \Gamma(r + \alpha + 1).	
\end{equation}
According to Table 1.1 of 
\cite{Gautschi2004}, $\alpha_k = 2 k + \alpha + 1$ for $k = 0, 1, 2, \dots$,
$\beta_0 = \Gamma(1 + \alpha)$ and $\beta_k = k (k + \alpha)$ for $k = 1, 2, 3, \dots$.

For $\alpha = 0$, the Laguerre case, the maximum of the absolute values of the differences between the computed recursion coefficients and these known results was approximately $3.9 \times 10^{-121}$.  

For $\alpha = 1$, the maximum of the absolute values of the differences between the computed recursion coefficients and these known results was approximately $3.9 \times 10^{-121}$.

\subsection{Computation of some classical Gauss quadrature nodes and weights}

We check our \texttt{R} code for all of the steps described in Appendix A, together with the validity of our rough heuristic (described in Section 2) for the assessment of the precision of our computed 
high-precision Gauss quadrature nodes and weights by comparing these computed values with known results.
The known results were obtained using the online calculator \cite{Keisan_Casio_Online_Calculator_2022}, with 50 decimal digits of precision. We consider Gauss Hermite, Gauss Legendre and Generalized Gauss Laguerre quadrature with number of nodes $n = 4$ and 
$n = 16$. In every case, the values of 
$\big(d_{\tau}(2), d_{\tau}(3), d_{\tau}(4), d_{\tau}(5)\big)$,
$\big(d_{\lambda}(2),d_{\lambda}(3), d_{\lambda}(4), d_{\lambda}(5)\big)$, $L_{\text{nodes}}$ and
$L_{\text{weights}}$ 
provided us with some reasonable degree of assurance that 
$\texttt{b}_5$ bits of precision is sufficiently large that, after
the high-precision nodes and weights are converted to double precision using  
the \texttt{asNumeric} function from the \texttt{R} package \texttt{Rmpfr}, the maximum possible precision is achieved. 

\subsubsection{Gauss Hermite quadrature}

For $n = 4$ our code produced high-precision results in full agreement with the 50 decimal digits of precision provided by \cite{Keisan_Casio_Online_Calculator_2022}. For $n = 16$ our code produced high-precision results in full agreement with the 50 decimal digits of precision provided by \cite{Keisan_Casio_Online_Calculator_2022}, except for one node which agreed to only 49 decimal digits of precision. 

\subsubsection{Gauss Legendre quadrature}

For both $n = 4$ and $n = 16$, our code produced high-precision results in full agreement with the 50 decimal digits of precision provided by \cite{Keisan_Casio_Online_Calculator_2022}. 

\subsubsection{Generalized Gauss Laguerre quadrature}

Suppose that $\alpha = 0$, the Laguerre case. 
For $n = 4$ our code produced high-precision results in full agreement with the 50 decimal digits of precision provided by \cite{Keisan_Casio_Online_Calculator_2022}.  For $n = 16$ our code produced high-precision results in full agreement with the 50 decimal digits of precision provided by \cite{Keisan_Casio_Online_Calculator_2022}, except for one node which agreed to only 49 decimal digits of precision.

Suppose that $\alpha = 1$.
For both $n = 4$ and $n = 16$, our code produced high-precision results in full agreement with the 50 decimal digits of precision provided by \cite{Keisan_Casio_Online_Calculator_2022}. 

%

\section{Discussion}

Our \texttt{R} package \texttt{custom.gauss.quad} is freely available. The algorithms used are very simple and easy to understand.
The computation of
 the recursion coefficients uses moment determinants as described in Theorem 2.2 of \cite{Gautschi2004}. These recursion coefficients are then used in the three-term recurrence relation as described in Theorem 1.27 of \cite{Gautschi1968}. The Gauss quadrature nodes are found by straightforward root finding and the corresponding weights are computed using the first displayed equation on p.23 of \cite{Gautschi2004}. 
 
Our code also includes a procedure for the provision of some reasonable degree of assurance that 
the largest number of bits of precision used is sufficiently large that, after
the nodes and weights are converted to double precision using  
the \texttt{asNumeric} function from the \texttt{R} package \texttt{Rmpfr}, the maximum possible precision (in double precision) is achieved.

%

\newpage

\appendix

\section*{Appendix A: Implementation of the moment-based method via moment determinants}

Steps 1 to 4, described below, are carried out by our \texttt{R} code to compute the custom-made Gauss quadrature nodes $\tau_1, \dots, \tau_n$
and weights $\lambda_1, \dots, \lambda_n$,
where the first step is a moment-based method using moment determinants.
Suppose that 
$n$ is given.
This code 
uses the high-precision arithmetic, with a specified number of bits of precision \texttt{b}, provided by the \texttt{R} package \texttt{Rmpfr} throughout. This code is applicable for whenever there is a
formula for the moments
in terms of the mathematical functions that can be computed to arbitrary precision using this package.

The nodes and weights may be expressed in terms of the eigenvalues and eigenvectors of the Jacobi matrix (see e.g. Theorem 1.3.1 of \citeauthor{Gautschi2004},  \citeyear{Gautschi2004}). However, the \texttt{Rmpfr}
package does not provide high-precision computation of eigenvalues and eigenvectors of a matrix. Consequently, we have computed the nodes and weights using a different approach. 
We first compute the nodes, by finding the roots of 
$\pi_n(\, \cdot \, )$
the monic orthogonal polynomial of degree $n$ with respect to the measure
specified by the function $f$. 
In our code, we represent a polynomial by the vector of its coefficients. 
The weights could be computed using either formula
(2.7.8) or formula (2.7.5.9) of \cite{DavisRabinowitz1984},
which are expressed in terms of
orthonormal polynomials with respect to this measure.
We avoid the need for orthonormalization by using the first displayed equation on p.23 of \cite{Gautschi2004} to compute the weights. 

\medskip

\noindent \textbf{Step 1}: Using the formulas, in terms of determinants of matrices whose entries are moments, given by Theorem 2.2 on p.54 of 
\cite{Gautschi2004}, compute the recursion coefficients
$\alpha_0, \alpha_1, \dots, \alpha_{n-1}$ and 
$\beta_1, \beta_2, \dots, \beta_{n-1}$. 
The determinant of a square matrix with more than 2 rows is computed using the LU decomposition. The code for this decomposition was obtained by translating the function \texttt{lu.description} from the \texttt{R} package \texttt{matrixcalc} (\citeauthor{matrixcalc},
\citeyear{matrixcalc}) into a form that uses the high-precision arithmetic provided by the \texttt{R} package \texttt{Rmpfr}. 

\medskip

\noindent \textbf{Step 2}: Using the three-term recurrence relation for monic orthogonal polynomials given by Theorem 1.27 on p.10 of \cite{Gautschi2004}, compute $\pi_n( \, \cdot \,)$, the monic orthogonal polynomial of degree $n$ with respect to the measure specified by the function $f$. 

\medskip

\noindent \textbf{Step 3}: The nodes $\tau_1, \dots, \tau_n$ are roots of $\pi_n( \, \cdot \,)$. Since $\pi_n( \, \cdot \,)$ is a real polynomial with real roots, we first compute Laguerre's bounds on these roots (\citeauthor{Laguerre1880}, \citeyear{Laguerre1880}), described e.g. by \cite{JensenStyan1999}. By Theorem 1.46 of \cite{Gautschi2004}, these roots should also 
belong to the support of the measure specified by the function $f$.
 Then the interval between the lower and upper bounds on the roots is evenly divided into $100n$ intervals. We then isolate the roots using the changes in sign of evaluations of the polynomial over the successive $100n + 1$ endpoints of these intervals. Finally, the roots are calculated using the function 
 \texttt{unirootR} from the \texttt{R} package \texttt{Rmpfr}. As stated in documentation for this package: `\texttt{unirootR()} is a  
 ``clone'' of \texttt{uniroot()}, written entirely in \texttt{R}, in a way that it works with \texttt{mpfr}-numbers as well.' The 
 \texttt{R} function \texttt{uniroot}  implements Brent's method for finding a zero of a function (see Chapter 4 of \citeauthor{Brent1973},
\citeyear{Brent1973}).

\medskip

\noindent \textbf{Step 4}: We compute the weight $\lambda_j$  as follows ($j = 1, 2, \dots, n$). First compute the Lagrange interpolation polynomial 
\begin{equation}
\ell_j(x) = \prod_{i=1}^{n} \frac{x - \tau_i}{\tau_j - \tau_i},
\end{equation}
which we represent by the $n$-vector of its coefficients. We then compute 
\begin{align*}
	q_j(x) &= \ell_j^2(x) 
	\\
	&= a_0 + a_1 x + \dots + a_{2(n-1)} x^{2(n-1)},
\end{align*}
say, which we represent by the $(2n-1)$-vector of its coefficients. By the first displayed equation on p.23 of \cite{Gautschi2004},
\begin{align*}
\lambda_j &= \int_{-\infty}^{\infty} q_j(x) \, f(x) \, dx
\\
&= a_0 + a_1 \, \int_{-\infty}^{\infty} x \, f(x) \, dx + \dots + a_{2(n-1)} \, \int_{-\infty}^{\infty} x^{2(n-1)}  \, f(x) \, dx.
\end{align*}

\section*{Appendix B: The inputs and output of the \texttt{R} \newline function \texttt{moments}}

The \texttt{R} function \texttt{moments} has the following inputs.

\begin{description}
	
	\item[\texttt{which.f}] 
	
\ a list specifying the nonnegative weight function $f$, with the following three components: (i) name (in the form of a character string), (ii) support specified by a 2-vector of the endpoints of the interval, (iii) parameter vector when $f$ belongs to a family of weight functions and is specified by the value of this parameter vector (if $f$ is already fully specified then the parameter vector is set to \texttt{NULL}).	

\item [\texttt{r}]

\ a nonnegative integer, specifying that it is the $r$'th moment for the weight function $f$ that is to be computed.

\item[\texttt{nbits}]

\ the number of bits in the multiple precision numbers used by the \texttt{R} package \texttt{Rmpfr} to carry out the computation of the $r$'th moment.

\end{description}

The \texttt{R} function \texttt{moments} has code segments for the following four weight functions $f$:

\smallskip

\noindent \textbf{(1) scaled chi pdf} 

\smallskip

\noindent This weight function, given by  \eqref{eqn_pdf_of_Example}, is specified by first assigning the value of $m$ and then using the following \texttt{R} command:

\begin{verbatim}
which.f <- list(name="scaled.chi.pdf", support=c(0, Inf), 
                parameters=m)
\end{verbatim}

\noindent For this weight function, the \texttt{R} function \texttt{moments} includes 
 the following segment of code for the computation of the $r$'th moment,  given by \eqref{eqn_moments_scaled_chi_pdf}:

\begin{verbatim}
if (which.f$name == "scaled.chi.pdf"){
    m <- which.f$parameters
    if (r == 0){
        return(mpfr(1, nbits))
    }
    mp.2 <- mpfr(2, nbits)
    mp.r <- mpfr(r, nbits)
    mp.m <- mpfr(m, nbits)
    term1 <- (mp.r/ mp.2) * log(mp.2 / mp.m)
    term2 <- lgamma((mp.r + mp.m) / mp.2)
    term3 <- lgamma(mp.m / mp.2)
    return(exp(term1 + term2 - term3))
}
\end{verbatim}

\smallskip

\noindent \textbf{(2) Hermite} 

\smallskip

\noindent This weight function, given by \eqref{eqn_Hermite_weight_function},
is specified by the following \texttt{R} command:

\begin{verbatim}
which.f <- list(name="Hermite", support=c(-Inf, Inf), 
                parameters=NULL)
\end{verbatim}

\noindent For this weight function, the \texttt{R} function \texttt{moments} includes 
the following segment of code for the computation of the $r$'th moment,  given by \eqref{eqn_moments_Hermite}	:

\begin{verbatim}
if (which.f$name == "Hermite"){
    if (r == 0){
        pi.mp <- Const("pi", nbits)
        return(sqrt(pi.mp))
    }
    if (2 * as.integer(r/2) != r){
        return(mpfr(0, nbits))
    }
    num.mp <- mpfr(r + 1, nbits)
    denom.mp <- mpfr(2, nbits)
    return(gamma(num.mp / denom.mp))
}
\end{verbatim}

\smallskip

\noindent \textbf{(3) Legendre} 

\smallskip

\noindent This weight function,
given by \eqref{eqn_Legendre_weight_function},
is specified by the following \texttt{R} command:

\begin{verbatim}
which.f <- list(name="Legendre", support=c(-1, 1), parameters=NULL)
\end{verbatim}

\noindent For this weight function, the \texttt{R} function \texttt{moments} includes 
the following segment of code for the computation of the $r$'th moment,  given by \eqref{eqn_moments_Legendre}:

\begin{verbatim}
if (which.f$name == "Legendre"){
  if (2 * as.integer(r/2) != r){
        return(mpfr(0, nbits))
    }
    num.mp <- mpfr(2, nbits)
    denom.mp <- mpfr(r + 1, nbits)
    return(num.mp / denom.mp)
}
\end{verbatim}

\smallskip

\noindent \textbf{(4) Generalized Laguerre} 

\smallskip

\noindent This weight function,
given by \eqref{eqn_Generalized_Laguerre_weight_function},
is specified by first assigning the value of $\alpha$, coded as \texttt{alpha.GGL}, and then using the following \texttt{R} command:

\begin{verbatim}
which.f <- list(name="Generalized.Laguerre", support=c(0, Inf),
                parameters=c(alpha.GGL))
\end{verbatim}

\noindent For this weight function, the \texttt{R} function \texttt{moments} includes 
the following segment of code for the computation of the $r$'th moment,  given by \eqref{eqn_moments_Generalised_Laguerre}:

\begin{verbatim}
if (which.f$name == "Generalized.Laguerre"){
    alpha.GGL <- which.f$parameters
    term.mp <- mpfr(r + alpha.GGL + 1, nbits)
    return(gamma(term.mp))
}
\end{verbatim}

\noindent The output of the \texttt{R} function \texttt{moments} is the 
$r$'th moment with number of bits of precision \texttt{nbits} used in its computation, via the \texttt{R}  package \texttt{Rmpfr}.

\section*{Appendix C: The inputs and outputs of the \texttt{R} \newline function \texttt{custom}}

The inputs to the \texttt{R} function \texttt{custom} are 
\texttt{which.f}, described in detail in Appendix B, and \texttt{n} the 
number of Gauss quadrature nodes.
The outputs of this function consist of a printout, which includes the vector $(\texttt{b}_1, \dots, \texttt{b}_5)$
of the number of bits of precision used in the computations
described in Section \ref{sect_moment_based_method}, and a list
whose elements have the following names:

\begin{itemize}[leftmargin=*]
	
	\item 
	
	\texttt{list.Gauss.nodes} is a list with the following 5 elements: 
	\newline
	\phantom{1} \qquad $\texttt{list.Gauss.nodes[[1]]}, \dots, \texttt{list.Gauss.nodes[[5]]}$, 
	\newline
	which are the $n$ Gauss quadrature nodes 
	computed using numbers of bits of precision $\texttt{b}_1, \dots, \texttt{b}_5$,
	respectively. 
	
		\item 
		
\texttt{list.Gauss.weights} is a list with the following 5 elements:
\newline
\phantom{1} \qquad  $\texttt{list.Gauss.weights[[1]]}, \dots, \texttt{list.Gauss.weights[[5]]}$, 
\newline
which are the $n$ Gauss quadrature weights 
computed using numbers of bits of precision $\texttt{b}_1, \dots, \texttt{b}_5$,
respectively. 

\item

\texttt{mat.timings} is a $3 \times 5$ matrix, where the columns correspond to the elements of the vector 
and the rows are the computation times (in seconds) for Steps 1 and 2 (taken together), Step 3 and Step 4 described in Appendix A.

\item 

\texttt{max.abs.diffs.nodes} is the vector 
$\big(d_{\tau}(2), d_{\tau}(3), d_{\tau}(4), d_{\tau}(5)\big)$.

\item 

\texttt{sum.abs.diffs.weights} is the vector 
$\big(d_{\lambda}(2),d_{\lambda}(3), d_{\lambda}(4), d_{\lambda}(5)\big)$.
	
\item 

\texttt{L.nodes} and \texttt{L.weights}, which are $L_{\text{nodes}}$ and $L_{\text{weights}}$, respectively. 

\item

\texttt{asNumeric.nodes} and \texttt{asNumeric.weights} are the 
$n$ double precision Gauss quadrature nodes and weights, respectively,  obtained by applying 
the \texttt{asNumeric} function from the \texttt{R} package \texttt{Rmpfr}
to the most accurate approximations to the Gauss quadrature nodes and
weights, namely,
\texttt{list.Gauss.nodes[[5]]} and \newline \texttt{list.Gauss.weights[[5]]}.

\end{itemize}

\end{document}